\documentclass[sigconf]{acmart}

\AtBeginDocument{%
  }

\copyrightyear{2024}
\acmYear{2024}
\setcopyright{rightsretained}
\acmConference[SCORED '24]{Proceedings of the 2024 Workshop on Software Supply Chain Offensive Research and Ecosystem Defenses }{October 14--18, 2024}{Salt Lake City, UT, USA}
\acmBooktitle{Proceedings of the 2024 Workshop on Software Supply Chain Offensive Research and Ecosystem Defenses (SCORED '24), October 14--18, 2024, Salt Lake City, UT, USA}
\acmDOI{10.1145/3689944.3696166}
\acmISBN{979-8-4007-1240-1/24/10}

\usepackage{xcolor}
\usepackage{colortbl}
\usepackage{float}
\usepackage[utf8]{inputenc}
\usepackage[T1]{fontenc}
\usepackage{amsthm,amsfonts}
\usepackage{tikz,lmodern}
\usepackage[most]{tcolorbox}
\usepackage{listings}
\usepackage{go/lang}  
\usepackage{go/style} 
\usepackage{go/json} 
\usepackage{booktabs}
\usepackage{hyperref}
\usepackage{svg}

\definecolor{lightgray}{gray}{0.8}
\newcommand{\toolname}[1]{\emph{GoSurf}}

\definecolor{lightgray}{rgb}{0.9, 0.9, 0.9}

\newtheoremstyle{styleth}%
{10pt}
{3pt}
{\color{black}}
{}
{\bfseries\color{black}}
{}
{.5em}
{}
\theoremstyle{styleth}
\newtheorem*{thm}{Definition}

\newcommand{\statetheoremsolid}[2]{
  \par\noindent\tikzstyle{mybox} = [draw=black,fill=blue!5,
   thick,rectangle,rounded corners,inner sep=6pt]
  \begin{tikzpicture}
    \node [mybox] (box){%
    \begin{minipage}{#1}{#2}\end{minipage}
   };
  \end{tikzpicture}
}

\begin{document}
\title[]{GoSurf: Identifying Software Supply Chain Attack Vectors in Go}

\author{Carmine Cesarano}
\email{carmine.cesarano2@unina.it}
\affiliation{
  \institution{Università degli Studi di Napoli Federico II}
  \city{Naples}
  \country{Italy}
}

\author{Vivi Andersson}
\email{vivia@kth.se}
\affiliation{
  \institution{KTH Royal Institute of Technology}
  \city{Stockholm}
  \country{Sweden}
}

\author{Roberto Natella}
\email{roberto.natella@unina.it}
\affiliation{
  \institution{Università degli Studi di Napoli Federico II}
  \city{Naples}
  \country{Italy}
}

\author{Martin Monperrus}
\email{monperrus@kth.se}
\affiliation{
  \institution{KTH Royal Institute of Technology}
  \city{Stockholm}
  \country{Sweden}
}

\begin{abstract}
In Go, the widespread adoption of open-source software has led to a flourishing ecosystem of third-party dependencies, which are often integrated into critical systems. However, the reuse of dependencies introduces significant supply chain security risks, as a single compromised package can have cascading impacts. Existing supply chain attack taxonomies overlook language-specific features that can be exploited by attackers to hide malicious code. In this paper, we propose a novel taxonomy of 12 distinct attack vectors tailored for the Go language and its package lifecycle. Our taxonomy identifies patterns in which language-specific Go features, intended for benign purposes, can be misused to propagate malicious code stealthily through supply chains. Additionally, we introduce \toolname{}, a static analysis tool that analyzes the attack surface of Go packages according to our proposed taxonomy. We evaluate \toolname{} on a corpus of 500 widely used, real-world Go packages. Our work provides preliminary insights for securing the open-source software supply chain within the Go ecosystem, allowing developers and security analysts to prioritize code audit efforts and uncover hidden malicious behaviors.  
\end{abstract}




\maketitle

\section{Introduction}
\label{sec:introduction}

Software development today heavily relies on open-source software (OSS) dependencies, leveraging the benefits of code reusability. These dependencies enable developers to reuse pre-written, tested, and community-vetted code, reducing development time and effort. This phenomenon is observed in all software ecosystems, including Java, Python and Go.

Despite its benefits, OSS also introduces security risks. A single compromised dependency in the software supply chain can put numerous downstream applications and services at risk. In recent years, there has been a growing trend of such supply chain attacks \cite{synopsis_report} where malicious actors compromise OSS projects to propagate malicious code to downstream applications. Real-world incidents, such as \textit{npm} typo-squatted packages \cite{npm_typosquatting}, malicious crypto miners uploaded to the Python Package Index (PyPI) \cite{sonatype_pypi_2024}, and potential repo-jacking attacks affecting more than 15K Go repositories \cite{repojacking_go}, highlight the harmful impact of software supply chain attacks across programming languages.

The Go programming language, known for its simplicity, concurrency, and performance, is not immune to supply chain attacks. Its supply chain demands specific attention due to its growing adoption in security-critical software like Kubernetes \cite{kubernetes} and Go Ethereum \cite{go-ethereum}. 
Indeed, there have already been reports of typosquatted Go packages, such as the widely used \texttt{'cli'} package, compromised in a malicious fork that introduced a malicious \texttt{'init'} function to collect private system information and exfiltrate it to a remote host \cite{typosquatting_go}.

In this paper, we study how attackers can hide and execute arbitrary code in Go dependencies. This important dimension of software supply chain security has been overlooked so far. On the one hand, existing attack taxonomies \cite{ohm2020backstabber,ladisa2023sok,ladisa2023hitchhiker} are generic and lack language-specific insights. On the other hand, existing Go security tools \cite{govulncheck,goast, gosec, govet,capslock} only cover a narrow range of possible software supply chain attack vectors in Go.

To address these challenges, we propose a novel taxonomy categorizing 12 distinct attack vectors specifically tailored for the Go ecosystem. This is the first systematic enumeration of stealthy ways to conceal malicious code in Go dependencies for propagation downstream. We make this taxonomy actionable with \toolname{}, a static analyzer for Go to identify the presence of these attack vectors in Go open-source packages. 

We evaluate the applicability of the proposed taxonomy on a set of 500 popular Go modules by using \toolname{} to quantify occurrences of attack vectors. Our analysis reveals that all 12 defined attack vectors can be successfully identified in real-world, popular Go projects, and most of these projects exhibit a large attack surface. In addition, we show a practical use case of \toolname{} by analyzing and comparing different version releases of a Go module, demonstrating how the attack surface evolves over time.

\toolname{} can play a crucial role in securing the supply chain of open-source Go projects by scrutinizing the attack surface of dependencies before integration, and continuously monitoring dependency updates for new attack vectors. The high volume of occurrences identified by \toolname{} underscores its potential as a foundational step towards more sophisticated analysis techniques, such as automated analysis of sensitive operations invoked within identified attack patterns, and prioritization. 

The main contributions of this paper are:
\begin{itemize}
    \item A novel taxonomy with 12 attack vectors for code execution specifically tailored for the Go software stack, showcased by runnable proof of concepts.
    \item The design and implementation of \toolname{} for statically analyzing the attack surface of Go packages, following the proposed taxonomy.
    \item An analysis of open-source, real-world Go packages to evaluate \toolname{}. We publicly release \toolname{} and the evaluation scripts     \footnote{\url{https://github.com/chains-project/GoSurf}}.
    
\end{itemize}


\section{Background on Golang}
\label{sec:background}
The Go programming language \cite{go_lang_ref}, commonly known as Golang, was created by Google in 2007. It is a systems language that addresses shortcomings of other programming languages by blending their distinctive characteristics \cite{go_language_design}, such as the static typing and runtime efficiency of C, the readability and usability of Python, and the high-performance multiprocessing features of Erlang. From a security standpoint, Go offers features like garbage collection, bounds checking, and type safety, which help prevent common vulnerabilities such as buffer overflows and memory leaks. Since its public release, Go has gained widespread adoption \cite{stackoverflow_survey}.

In Go, software is organized into units called Go modules, which are collections of related Go packages that are versioned and distributed together \cite{go_modules}. A Go package consists of one or more source files within the same directory. Projects can depend on entire modules or only on single packages, according to the specific needs of their application. 

Open-source Go modules are supplied to dependents through a distributed package management system. By publishing a Go module to a version control system (VCS) such as GitHub, any developer can import it by referencing its repository hosting location \cite{go_modules}. An official Go proxy maintains mirrors for the repositories, ensuring that modules are reliably available in the Go package ecosystem.

\begin{figure}[h]
  \centering
  \includegraphics[width=1\linewidth]{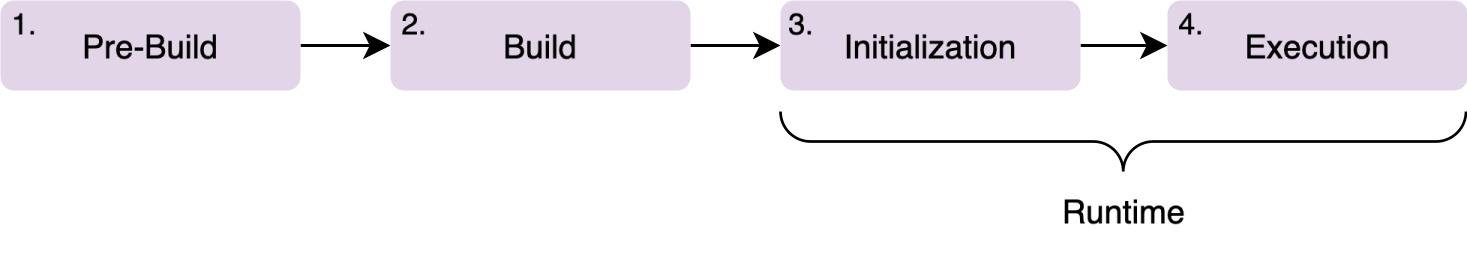}
  \caption{Go Package Lifecycle}
  \label{fig:lifecycle}
  \Description{Lifecycle for a Go Package}
\end{figure}

The Go module system and toolchain simplify dependency management throughout the phases shown in Figure \ref{fig:lifecycle}, and described in the following.

\vspace{-3mm}

\subsubsection*{1. Pre-Build}\label{subsubsec:pre-build} To create a Go module, developers can use the \verb|'go mod init'| command, which generates a manifest file named \verb|'go.mod'| (shown in Listing \ref{lst:go_mod}). This file lists all the required direct and indirect dependencies and ensures reproducible builds by pinning dependencies to specific versions \cite{go_modules}. During this phase, dependencies can be explicitly integrated into the Go module through the \verb|'go get'| command, which fetches and caches the required packages locally. After this, developers can test integrated packages (\verb|'go test'|) and utilize code generation tools (\verb|'go generate'|) to automate tasks like creating boilerplate code or generating environment-specific artifacts, before the build phase.

\begin{lstlisting}[language=go, style=go, caption={Example of a \texttt{go.mod} manifest file.}, label={lst:go_mod}, float=t]
module github.com/ethereum/go-ethereum
go 1.21
require (github.com/Azure/.../azblob v1.2.0
         github.com/Microsoft/go-winio v0.6.1 ...)
\end{lstlisting}

\subsubsection*{2. Build} During the compilation of a module, the \verb|'go build'| tool parses the manifest (\verb|'go.mod'|) and automatically fetches any required dependencies not cached locally. The Go compiler then compiles them into machine code \cite{go_compile}. The compiled code is linked with the Go runtime to create an executable binary. Go toolchain is explicitly designed to avoid code execution during this process, ensuring build-time security.

\subsubsection*{3. Initialization}\label{subsubsec:initialization} When executing the installed binary, the Go Runtime first initializes dependencies based on a dependency ordering \cite{go_lang_ref}. If package A imports package B, package B will be initialized before package A. First, package-level \textit{global variables} are initialized, in the order they are declared. Then, the Go runtime executes any package-level \href{https://go.dev/ref/spec#Package_initialization}{\texttt{init()}} functions. Global variables initialization and init functions can contain arbitrary code.

\subsubsection*{4. Execution}\label{subsubsec:execution} Finally, once all packages are initialized, the Go runtime will execute the \verb|'main'| function in the main package, as a \textit{goroutine} \cite{go_lang_ref}. A goroutine is a lightweight thread managed by the Go runtime, allowing for concurrent execution with lighter overhead than with traditional OS threads. The Go runtime also automatically manages memory allocation and garbage collection.


\section{Supply Chain Attack Vectors in Go}
\label{sec:attacks}

A supply chain attack typically involves two main stages: first, an attacker introduces malicious code in an upstream software component, which subsequently is propagated to downstream users. The second stage is the execution of this malicious code.
In this paper, we focus on supply chain attacks related to source code, and define `attack vector' as follows. 
\vspace{10pt}
\statetheoremsolid{0.45\textwidth}{
 \begin{thm}
   An attack vector is a language feature that can be potentially used by an attacker to hide and execute malicious code during the application lifecycle, at some point between build and production.
 \end{thm}
}
\vspace{-3pt}

Those source code attack vectors are highly programming-language specific.
We note that the usage of these features is not inherently malicious, and they are indeed mostly used for good. Yet, we claim that as potential attack vectors, they should be prioritized for security audits, as they are more prone to hide of execution of malicious code. Establishing a robust taxonomy of these attack vectors is critical in software supply chain security. Although previous efforts \cite{ladisa2023hitchhiker, ohm2020backstabber} have yielded general taxonomies for supply chain attack strategies for arbitrary code execution, they are not tailored to the unique characteristics and idiosyncrasies of specific programming languages and stacks. 

In this section, we propose a novel language-specific taxonomy to classify attack vectors for code execution in Go packages. With this taxonomy, we can help researchers and security experts to effectively understand and identify the unique risks linked to Go dependencies. To develop a Go-specific taxonomy of supply chain attack vectors, we began by collecting two existing Go-specific vectors from the literature \cite{ladisa2023hitchhiker} (constructors and init functions). Next, we reviewed known attack vectors in other languages \cite{ohm2020backstabber} and their relevance in Go, identifying an additional applicable vector (testing functions). Finally, by analyzing the Go documentation, we identified nine other  vectors relying on Go-specific features, not covered by any previous work. To validate these findings, we curated a dataset of executable proof-of-concept attacks written in Go, included in the project repository.

We categorized the identified attack vectors according to three phases of the package lifecycle: pre-build, initialization, and execution (see Sec. \ref{sec:background} for a detailed explanation of these phases). For each of the identified attack vectors, we provide a detailed explanation of the Go-specific feature and its intended purpose as well as how it can be exploited to perform an attack on the supply chain.

\vspace{-5pt}
\subsection{Malicious Code at Pre-Build Time}
\label{subsec:pre-build}
During the pre-build phase, third-party dependency integration encompasses various operations. The Go toolchain provides two primary tools to streamline these operations: \verb|'go generate'| for automating code generation tasks, and \verb|'go test'| for testing the code. Attackers can exploit both operations to hide and execute malicious code. 

\subsubsection*{P1. Static Code Generation}
\label{subsubsec:go_generate}
The \href{https://pkg.go.dev/cmd/go/internal/generate}{\texttt{go generate} tool} enables developers to define and execute \textit{code generators}, which run commands and generate Go source code. Go generators are defined in comments using the \verb|'//go:generate'| directive, followed by a shell command to be executed, e.g. \verb|'//go:generate cmd'|. Then, when the \verb|'go generate'| tool is run, source files in the package are scanned for these directives, executing the specified shell commands. It is important to note that \verb|'go generate'| is not part of the \verb|'go build'| command by default and must be explicitly run before building the project \cite{go_generate_pike}.

\paragraph{Attack} This feature poses two main security concerns. First, generators can execute arbitrary shell commands, allowing attackers to insert malicious code into the directive to be directly executed. Second, generators can generate additional source code on the fly to attach to a Go project before building, thereby enabling the injection of malicious code into the final executable during the subsequent build.

Malicious code hidden in these directives can be unintentionally executed by a developer. This can happen when imported dependencies include code generation operations that need to be run before compilation. 

In addition, CI/CD pipelines might automatically run the \verb|'go generate'| command when invoked from a build script. For example, as evidenced by real-life Go projects such as Istio \cite{istio_generators}, Terraform \cite{terraform_generators}, and Vault \cite{vault_generators}, it is a common practice to run code generation in Makefile scripts.

A real-world example \cite{CasaOS} of this pattern is shown in Listing \ref{lst:generate}. Although the directive does not directly execute any harmful commands, the code fetched from an external source can be inherently malicious.

\begin{lstlisting}[language=go,style=go,caption={Usage of the \texttt{go:generate} directive.},label={lst:generate}]
//go:generate bash -c "mkdir -p codegen && go run github.com/deepmap/oapi-codegen/cmd/oapi-codegen@v1.12.4 -generate types,server,spec -package codegen api/casaos/openapi.yaml > codegen/casaos_api.go"
\end{lstlisting}

\subsubsection*{P2. Testing Functions}
\label{subsubsec:testing}
The Go Toolchain includes built-in features to facilitate robust and efficient software testing. Two central components of the Go testing framework are the \verb|'go test'| tool and the standard library \href{https://pkg.go.dev/testing}{\texttt{testing}}, which automates the execution of test cases. Developers can create test files ending in \verb|'_test.go'|, and define functions starting with the prefix \verb|'Example'|, \verb|'Test'|, \verb|'Benchmark'| or \verb|'Fuzz'| to design examples, tests, benchmarks, and fuzzing, respectively. These functions are automatically executed by the \verb|'go test'| tool. This feature is intended to be used during the testing stage of development or as part of the CI actions. Dependency consumers may also use it to test imported code before integration and compilation.

\paragraph{Attack} 
Test suites create a low-visibility environment with potentially large amounts of code, allowing attackers to hide malicious code for seamless execution during build integration. This code might compromise either the developer machine or the CI runners. Testing functions may be overlooked in security reviews because of the absence of business logic that ends up in production. However, some testing functions can be triggered during post-deployment smoke tests. This provides attackers with the opportunity to execute arbitrary code on production machines.

\subsection{Malicious Code at Initialization Time}
\label{subsec:initialization}
As mentioned in Sec. \ref{sec:background}, the Go Runtime supports initialization operations executed before the main logic by means of \textit{global variable initialization} and \verb|'init'| functions. Although these mechanisms are useful for initializing packages, arbitrary code can be injected and executed through them.

\subsubsection*{I1. Global Variable Initialization}
\label{subsubsec:global_variable}
In Go, any variable declared outside a function is referred to as a \textit{package-level global variable}. Global variables can be initialized using simple expressions, such as a composite literal, or when more complex operations are required by the return value of regular or anonymous function calls. From a runtime perspective, functions (regular or anonymous) invoked on the right-hand side of a global variable declaration statement, are executed promptly during package initialization, before the importing program executes both the \verb|'main'| and any \verb|'init'| functions.

\paragraph{Attack}
The initialization logic creates a stealthy attack vector. Because global variable initializations are executed before the \verb|'main'| function, an attacker can establish a foothold and potentially manipulate the application's behavior from the outset. Furthermore, this initialization logic is executed silently whenever a compromised package is imported, either directly or transitively, through nested dependencies, even if the package is not used later. Examples of both regular and anonymous functions used for global variable initialization are presented in Listing \ref{lst:global_var}.

\begin{lstlisting}[language=go,style=go,caption={Global variable initialization with function calls.},label={lst:global_var}]
func UnsafeMethod() string {
	// malicious code
}
var var1 string = unsafeMethod()
var var2 string = func() string {
    //malicious code
}() 
\end{lstlisting}

\subsubsection*{I2. Initialization Hooks}
\label{subsubsec:init}
Initialization tasks can also be defined in \verb|'init()'| functions. After initializing all global variables in a package, the Go runtime runs the \verb|'init()'| functions of dependencies. 

\paragraph{Attack} 
As for global variable initialization, an attacker can inject malicious code into an \verb|'init'| function to execute their payload automatically during the import process, before the main program starts. Note that Go removes unused dependencies such that if a package is imported but never directly used, its \verb|'init()'| function will not be executed. However, when importing a package with an underscore prefix, Go prevents the dependency from being removed, which ensures its \verb|'init'| function always runs \cite{effective-go}.

\subsection{Malicious Code at Execution Time}
\label{subsec:execution-time}
In Go, certain constructs can mask and execute harmful code during the main execution phase. These include hidden dynamic behavior and unsafe features.

\subsubsection*{E1. Constructor Methods}
\label{subsubsec:constructor}
In Go, developers define custom functions that serve as constructors for structures following a naming convention such as \verb|'NewStructName'| \cite{go_naming_conventions}. If the variable being initialized is the primary type for the given package, the constructor can be named \verb|'New'| without suffix. 
 
\paragraph{Attack} Constructors are called to create a new instance of a structure, and any malicious code embedded into constructors will execute during each instantiation. Since constructors are often expected to only handle simple initialization tasks, they may be overlooked as potential vectors for malicious activity. This makes hiding malicious code in constructors an effective method for attackers to compromise a system repeatedly.

\subsubsection*{E2. Reflection}
\label{subsubsec:reflection}
Reflection in Go enables dynamic inspection and runtime modification of types and methods. This is accomplished through the \href{https://pkg.go.dev/reflect}{\texttt{reflect} package}. When a method is invoked through reflection, the actual method called is determined at runtime rather than at compile-time. Developers use reflection to build generic libraries, implement serialization and deserialization mechanisms, and create flexible APIs that can operate on various types without needing to know their specifics at compile-time \cite{go_reflection}. 

\paragraph{Attack} An attacker can exploit reflection to dynamically inject and execute arbitrary code at runtime. For instance, consider the scenario shown in Listing \ref{lst:reflection}; reflection is used to call \verb|'UnsafeMethod'| dynamically. The method name to be called is stored in a variable that an attacker can manipulate (lines 10-11) to execute arbitrary methods. This ability to invoke methods dynamically makes it easier for attackers to hide malicious payloads and evade static security analysis \cite{sayar2023depth}. 

\begin{lstlisting}[language=go,style=go,caption={Indirect method invocation through reflection.},label={lst:reflection}]
type Mytype string
func (t MyType) Safe() {
	// benign code
}
func (t MyType) Unsafe() {
	// malicious code
}
func main() {
    var target MyType
    var methodName string = "Safe"
    methodName = "Unsafe" // hidden manipulation
    v := reflect.ValueOf(target)
    m := v.MethodByName(methodName)
    m.Call(nil)
\end{lstlisting}

\subsubsection*{E3. Interface Polymorphism}
\label{subsubsec:interfaces}
In Go, polymorphism is achieved through \href{go_interfaces}{\texttt{interfaces}}, which specify a set of method signatures, i.e., behaviors. Interfaces in Go are structurally typed rather than nominally typed. Any type is considered to implement an interface if it is structurally equivalent to the interface (i.e., implements all its methods) regardless of the explicit declaration. When an interface is used, the actual invoked method is dynamically dispatched based on the method set of the underlying type. Common use cases include dependency injection, creating mocks for unit testing, handling events, and defining callbacks. 

\paragraph{Attack} Interfaces in Go can be exploited to achieve polymorphic behavior when malicious code replaces a benign implementation at runtime. Listing \ref{lst:interfaces} shows a scenario where an interface designed to call \verb|'Execute'| on a safe type is subverted. By converting the instance of \verb|'SafeType'| with \verb|'UnsafeType'|(lines 15-16), the attacker ensures that the malicious implementation of \verb|'Execute'| is invoked (line 17). The insidious nature of this attack lies in the ability to hide the malicious implementation behind an innocuous-looking interface. In addition, the subtle replacement can leverage dynamic or indirect assignment of types, potentially being concealed deep within the code, making it difficult to detect during code reviews or static analysis. 

\begin{lstlisting}[language=go,style=go,caption={Indirect method invocation through interfaces.},label={lst:interfaces}]
type SafeInterface interface {
	Execute()
}
type SafeType struct{}
func (b SafeType) Execute() {
    // benign implementation
}
type UnsafeType struct{}
func (m UnsafeType) Execute() {
    // malicious implementation
}
func main() {
    var safeVar SafeInterface = SafeType{}
    safeVar.Execute()       // Safe code exec
    var unsafeVar SafeInterface = UnsafeType{}
    safeVar = unsafeVar     // hidden conversion
    safeVar.Execute()       // Unsafe code exec
\end{lstlisting}

\subsubsection*{E4. Unsafe Pointers}
\label{subsubsec:unsafe}
In Go, using safe abstractions such as arrays, slices, and maps ensures built-in checks that prevent common memory violations such as buffer overflows, null pointer dereferences, and out-of-bound memory accesses. However, to perform low-level programming tasks or to optimize certain performance-sensitive applications, the \href{https://pkg.go.dev/unsafe}{\texttt{unsafe} package} can be utilized. This package allows for the definition of the type \verb|'unsafe.Pointer'|, which bypasses Go's safety checks. For example, developers can perform pointer conversions, pointer arithmetic, and interpret the memory layout of complex data structures in ways that would otherwise be restricted.  

\paragraph{Attack} Safe usage of \textit{unsafe pointers} is possible, however, several examples of patterns that introduce security risks exist \cite{costa2021breaking, lauinger2020uncovering}. First, attackers can exploit unsafe pointers to perform unauthorized memory operations. As shown in Listing \ref{lst:unsafe_execution}, unsafe pointers can be misused to create a function pointer variable and set its value to the address of an arbitrary function (lines 7-8), thereby effectively allowing the execution of any code at runtime.

Similarly, Listing \ref{lst:unsafe_memory} shows how unsafe pointers can be misused to access out-of-bound memory locations, potentially leading to information disclosure or memory corruption vulnerabilities. 

\begin{lstlisting}[language=go,style=go,caption={Unsafe Pointers for Arbitrary Execution.},label={lst:unsafe_execution}]
func targetFunction() {// malicious code}
func main() {
	type FuncType func()
	targetFunc := targetFunction
	var funcPtr FuncType
	funcPtr = *(*FuncType)(unsafe.Pointer(&targetFunc))
	funcPtr()
}
\end{lstlisting}

\begin{lstlisting}[language=go,style=go,caption={Unsafe Pointers for Memory Access.},label={lst:unsafe_memory}]
data := [4]int{1, 2, 3, 4}
ptr := unsafe.Pointer(uintptr(unsafe.Pointer(&data[0])) + unsafe.Sizeof(data[0])*4)
value := *(*int)(ptr)
fmt.Printf("Out-of-bounds value: %d\n", value)
\end{lstlisting}

\subsubsection*{E5. CGO Static Code Linking}
\label{subsubsec:static_linking} 
\href{https://pkg.go.dev/cmd/cgo}{\textit{CGO}} is a foreign function interface allowing for static linking of C code within Go packages. Developers can write C snippets in a preamble comment of a Go file, directly defining C functions or importing external C header files. Thus, by using a pseudo-package \verb|'C'|, the defined C functions can be directly invoked within the Go program (shown in Listing \ref{lst:cgo}). When the Go compiler parses an import of \verb|'C'|, it invokes the C compiler on the defined C functions and links the C and Go object files. The CGO feature is commonly used to leverage existing C libraries to avoid error-prone re-implementations and bring performance benefits in Go \cite{sorniotti2023go}. 

\paragraph{Attack} From a security standpoint, CGO introduces risks, mainly concerning memory safety. While Go has built-in protection \cite{go_language_design} for common memory management errors, such as buffer overflows, dangling pointers, and memory leaks, C lacks these safeguards. Linking the C code in Go packages can reintroduce these vulnerabilities, which can be exploited by attackers. For example, memory violations can be used to gain unauthorized access or control over a system. In addition, attackers can manipulate function pointers in C to redirect execution to malicious code segments, bypassing Go's safety mechanisms.

\begin{lstlisting}[language=go,style=go,caption={C code invocation using CGO.},label={lst:cgo}]
/* #include <stdio.h>
void cFunction() { 
    // malicious code here
} */
import "C"
func main() {
    C.cFunction()
}
\end{lstlisting}

\subsubsection*{E6. Assembly Static Code Linking}\label{subsubsec:assembly}
Go supports static linking of assembly code \cite{go_assembler_guide}. Developers can write package-level assembly functions using Go Assembly. When compiling, any assembly file is assembled into object files by the \href{https://pkg.go.dev/cmd/asm}{Go assembler}, and then linked into the Go objects. These functions can be used in Go files by defining a function stub with the same name as that specified in the assembly file. The assembly functions can then be called as any normal function in Go (shown in Listing \ref{lst:asm}). Assembly is commonly leveraged for code optimization or in security-critical domains such as cryptography \cite{pike_assembler}. 

\paragraph{Attack} The complex syntax of assembly allows attackers to embed malicious code that can be challenging to audit due to its low-level nature. A lack of analysis support for assembly and the requirement of specialized knowledge (specifically in the go-specific assembly syntax) can leave this code poorly or completely unaudited. 
The use of empty functions as an assembly definition further conceals its underlying implementation. This makes assembly code a potent method for hiding malicious code.

\begin{lstlisting}[language=go,style=go,caption={Invocation of package-level defined assembly code.},label={lst:asm}]
func AsmFunction() int
func main() {
	AsmFunction() // malicious function
}
\end{lstlisting}

\subsubsection*{E7. Dynamic Library Linking}
\label{subsubsec:dynamic_linking}
In Go, developers can use the \href{https://pkg.go.dev/plugin}{\texttt{plugin}} package for dynamic loading and executing of external shared libraries within the main process of a running program. A plugin is a Go package with exported functions and variables built using the \verb|'-buildmode=plugin'| option. The Go runtime resolves the function and variable symbols provided by the plugin dynamically, enabling direct function calls between the main program and plugin. The use of plugins eliminates the overhead of inter-process communication connected to e.g. file system operations.

\paragraph{Attack.} While providing flexibility, plugins also introduce security risks. Plugins allow for changing the main program's behavior without recompiling it. An attacker can replace a plugin with a malicious version as shown in Listing \ref{lst:plugin}. This allows an untrusted, potentially dangerous library to be loaded into the main process stealthily without the developer's knowledge.

\begin{lstlisting}[language=go,style=go,caption={Simplified example call to dynamic library function.}, label={lst:plugin}]
func main() {
    pluginPath := "./malicious_plugin.so"
    p := plugin.Open(pluginPath)
    sym := p.Lookup("pFunc") // runtime resolution
    fn := sym.((func()))
    fn() // malicious function
}
\end{lstlisting}

\subsubsection*{E8. Dynamic External Execution}
\label{subsubsec:dynamic_execution}
Dynamic external execution involves running an executable as a separate process. In Go. this can be achieved through the \href{https://pkg.go.dev/os}{\texttt{os}} and \href{https://pkg.go.dev/syscall}{\texttt{syscall}} packages, offering methods like '\texttt{exec.Command}' and '\texttt{exec.CommandContext}'. Additionally, lower-level functions such as '\texttt{syscall.ForkExec}', '\texttt{syscall.Exec}', and '\texttt{os.StartProcess}' provide more control over process execution. This feature is intended for directly execute an external binary or shell commands. 

\paragraph{Attack} Dynamic external execution poses security threats. For example, binaries executed through them may be opaque and limit human and static analyses, facilitating the hiding of malicious code.
In addition, the possibility of invoking arbitrary commands that are potentially constructed dynamically can lead to OS command injection. This also introduces risks for denial-of-service attacks through fork bombs, i.e. recursive replication of processes.

\subsection{Attack Vectors in the Malware Lifecycle}
Based on the specific characteristics of each attack vector that we define, they can map to different stages of malware execution. During the \textit{infection stage}, attack vectors categorized in the pre-build phase (Sec. \ref{subsec:pre-build}), like using \textit{generators} and \textit{testing functions}, can be exploited to download preliminary malicious scripts or code, exfiltrate sensitive system information, or install backdoors. Subsequently, during the \textit{persistence stage}, attack vectors exploitable at the initialization time (Sec. \ref{subsec:initialization}) like \verb|init()| functions and global variables initialization can be used to establish a persistent presence on the infected system, for example, setting some registry entries or creating new scheduled tasks. As these functions are invoked automatically early in program execution, such malicious actions can occur stealthily without the developer's explicit invocation. Finally, all attack vectors we categorized as execution time (Sec. \ref{subsec:execution-time}) can be exploited for the \textit{payload execution stage} of malware, where the primary malicious activities are carried out.


\section{GoSurf}
\label{sec:gosurf}
In this paper, we introduce \toolname{}, a novel tool that analyzes the attack surface of Go modules according to the taxonomy of attack vectors described in Sec.  \ref{sec:attacks}. It reports the usage of language-specific features and programming idioms that can be leveraged for the corresponding attack vectors. \toolname{} is meant to aid open-source dependency analysis and audit to find malicious code.

\subsection{Design}
\begin{figure}[h]
  \centering
  \includegraphics[width=1\linewidth]{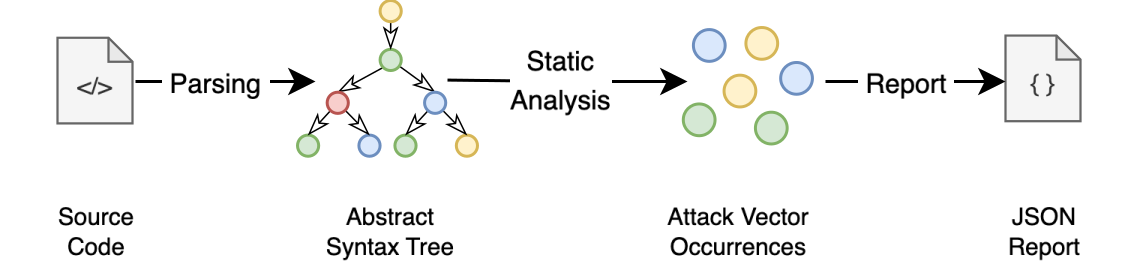}
  \caption{\toolname{} Tool Architecture}
  \label{fig:gosurf}
\end{figure}

\toolname{} follows a modular architecture consisting of three main parts: a parser, static analysis engine, and reporting component. The workflow is shown in Figure \ref{fig:gosurf}. 

The \textit{parser} component parses Go source code and constructs an Abstract Syntax Tree (AST). The \textit{static analysis engine} uses the AST and a set of rules to identify attack vectors based on the defined taxonomy. The \textit{reporting component} consolidates the analysis results presenting them in a user-friendly format. We implemented \toolname{} in Go, using 793 lines of code, and leveraging Go standard libraries. The details of each component are explained in the following.

\subsubsection{Parser}
The \toolname{} parser component enables analysis of Go modules, potentially composed of several packages. 

First, the parser collects all packages for analysis, starting with a root module path specified as a local directory. It recursively traverses all subdirectories to identify the module packages, recognizing them by the presence of \verb|.go| files containing valid package declarations (directive \verb|package|).

After gathering all packages from the root module path, the parser generates the abstract syntax tree (AST) representation of each Go file, using the Go standard libraries \href{https://pkg.go.dev/go/token}{\texttt{go/token}} and \href{https://pkg.go.dev/go/parser}{\texttt{go/parser}}.

\subsubsection{Static Analysis}
\toolname{} implements 12 analyzers to identify occurrences of every attack vector from our proposed taxonomy (Sec. \ref{sec:attacks}). The analyzers use the AST representation to inspect its nodes, primarily using the \verb|'ast.Inspect'| function from the \href{https://pkg.go.dev/go/ast}{\texttt{go/ast}} package. The analyzers can be divided into three groups based on their targeted nodes for inspection: an invocation, declaration, or comment. 

\paragraph{Invocation} The first group of analyzers (\textit{Constructor}, \textit{Interfaces}, \textit{Unsafe}, \textit{CGO}, \textit{Assembly}, \textit{Plugin}, \textit{Exec}) identify specific package and function invocation calls by inspecting \textit{call expressions} (\verb|'CallExpr'|). These analyzers search for function expressions belonging to an \textit{identifier} (\verb|'Ident'|) or \textit{selector expression} (\verb|'SelectorExpr'|). An identity is a standalone name, while a selector expression is a qualified identifier on the form \verb|'X.Sel'|, where \verb|'X'| is the package or receiver expression (e.g., a variable), and \verb|'Sel'| is the identifier being selected (e.g., a function name). For example, the \verb|Plugin| (Sec. \ref{subsubsec:dynamic_linking}) analyzer matches the AST node pattern of a call expression (\verb|'CallExpr'|) with a selector expression having the package name \verb|'plugin'|, selecting the method \verb|'Open'| (\verb|'plugin.Open()'|). 

Two invocation analyzers include a pre-processing step to identify the target method signature for inspection. The \verb|Interfaces| (Sec. \ref{subsubsec:interfaces}) analyzer collects all polymorphic methods by looking for functions with multiple receiver types. Conversely, the \verb|Assembly| (Sec. \ref{subsubsec:assembly}) analyzer collects all package-level assembly functions through regular expression matching in assembly files. Then, the invocation inspection technique is applied to these analyzers, identifying method calls to the previously collected method signatures.

\paragraph{Declaration} The second group of analyzers (\verb|GoTest|, \\ \verb|InitFunc|, \verb|GlobalVar|, \verb|Reflect|) analyze declarations in the AST. The \verb|GoTest| analyzer (Sec. \ref{subsubsec:testing}) matches \textit{function declaration} (\verb|'FuncDecl'|) nodes that have names starting with \verb|'Test'|, \verb|'Fuzz'|, \verb|'Benchmark'| or \verb|'Example'|. Similarly, the \verb|InitFunc| (Sec. \ref{subsubsec:init}) analyzer identifies function declarations named \verb|'init'|. The \verb|GlobalVar| (Sec. \ref{subsubsec:global_variable}) and \verb|Reflect| (Sec. \ref{subsubsec:reflection}) analyzers inspect \textit{generic declaration} (\verb|'GenDecl'|) nodes, looking for variable and constant declarations, and import declarations, respectively.

\paragraph{Comment} The third group inspects \textit{comments} in the AST. This technique concerns the \verb|GoGenerate| analyzer (Sec. \ref{subsubsec:go_generate}), which identifies generators in \verb|Comment| nodes with the prefix \verb|'//go:generate'|. 

\subsubsection{Reporting}
The reporting component of \toolname{} records and reports the occurrences of an attack vector. When encountering a match in an analyzer, the location information is logged, including package name, file path, and line number. Additional context-specific information is included, depending on the attack vector, such as the name of a method invocation. All occurrences are aggregated into JSON format and counted before presenting the results in the CLI. Listing \ref{lst:json} shows a \toolname{} report identifying usage of CGO and assembly.

\begin{lstlisting}[language=json, style=json, caption={Example JSON report.},label={lst:json}]
[{
    "PackageName": "issue27340",
    "Type": "cgo",
    "FilePath": ".../issue27340/a.go",
    "LineNumber": 41,
    "MethodInvoked": "C.issue27340CFunc"
    },
    {
    "PackageName": "main",
    "Type": "assembly",
    "FilePath": ".../testdata/main.go",
    "LineNumber": 8,
    "MethodInvoked": "linefrompc"
}]
\end{lstlisting}

\subsection{Usage of \toolname{}}
\toolname{} is a tool meant to be integrated into supply chain security processes. We outline two primary use cases.

\subsubsection{Pre-integration Auditing} 
\label{subsubsec:pre-integration}
Before integrating a third-party dependency, developers should conduct a thorough security audit. \toolname{} offers a comprehensive overview of the risks associated with a dependency and highlights those riskier areas of the codebase with a higher number of attack vectors. It also provides a valuable security metric for dependency consumers, helping them prioritize those dependencies with smaller attack surface.

\subsubsection{Version Update Monitoring}
\label{subsubsec:updates}
The attack surface of a software package is dynamic, with new attack vectors potentially introduced with each new version release. \toolname{} can be used to continuously monitor these changes, by comparing the attack surface of different versions of a package. This use case is particularly meaningful when a dependency with an initial trusted codebase is updated. Updates from untrusted authors with new attack vectors should be scrutinized. By using \toolname{}, upstream dependencies can be effectively monitored, ensuring consumers stay informed about risks.

To sum up, \toolname{} enhances the security posture in the Go ecosystem and identifies potential risks before actual software supply chain attacks.

\section{Evaluation}
\label{sec:evaluation}
In this section, we present initial experimental results obtained from analyzing \toolname{} on real-world Go projects. 

\subsection{Attack Surface in Real-World Go Modules}
\label{subsec:popular_modules}
In Sec. \ref{sec:attacks} we demonstrated the feasibility of the identified attack vectors, with executable proof of concepts. In this section, we conduct an experiment to evaluate the applicability of our proposed taxonomy, that is, that the 12 attack vectors can be successfully identified in real-world Go projects.

\paragraph{Methodology} 
To analyze the applicability of our taxonomy, we conducted a comprehensive analysis using \toolname{} across 500 Go modules. We selected these modules based on their number of dependents, specifically the top 500 most imported packages, indicating their relevance and popularity. This selection criteria ensures that we focus on modules with widespread real-world usage, which are natural targets of supply chain attacks. The selected projects cover diverse categories such as cloud orchestration, networking, blockchain, monitoring, and CLI applications, sourced from \href{https://libraries.io/}{\texttt{libraries.io}}. 

\begin{figure}[h]
  \centering
  \includegraphics[width=1\linewidth]{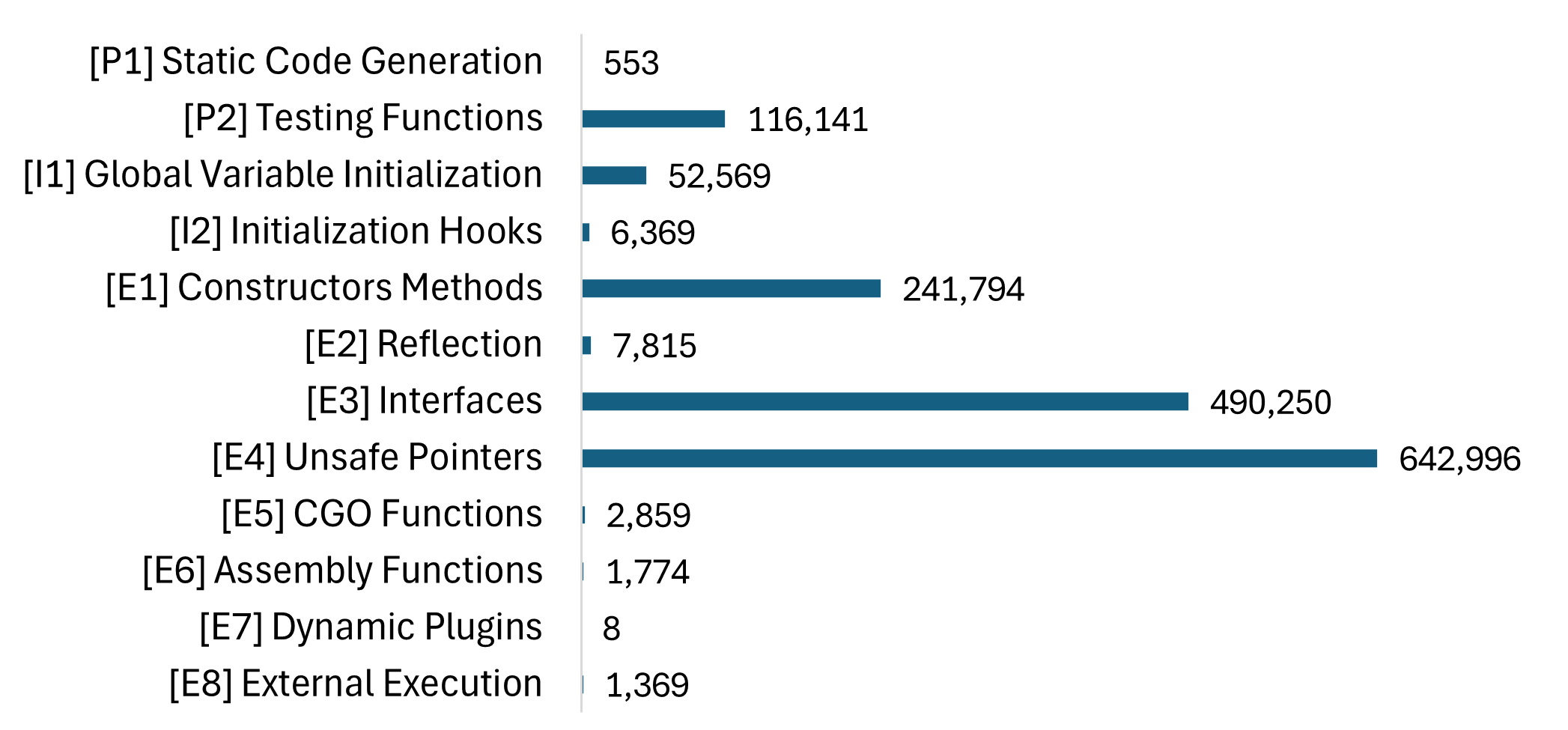}
  \caption{Go Attack Vector Prevalence in 500 Popular Go Projects. The rare ones are easier to forbid in order to secure a supply chain.}
  \label{fig:occurrences}
  \Description{graph for Attack Vectors Occurrences on top 500 Go Package}
\end{figure}

\paragraph{Results}
Using \toolname{}, we identified occurrences of some of the 12 attack vectors in each of the analyzed Go modules. Figure \ref{fig:occurrences} reports the total number of occurrences for each vector across the 500 analyzed modules.

The results indicate that \hyperref[subsubsec:testing]{P2} (testing functions), \hyperref[subsubsec:constructor]{E1} (constructors methods), \hyperref[subsubsec:interfaces]{E3} (interfaces) and \hyperref[subsubsec:unsafe]{E4} (unsafe) are the most frequently occurring attack vectors. Although their high prevalence may be connected to common benign Go programming practices, their usage call for additional auditing in a security-sensitive context. Overall, Table \ref{fig:occurrences} shows that malicious actors have ample room to choose where to hide malicious code execution, as argued in Sec. \ref{sec:attacks}.

Furthermore, the widespread usage of cross-language features such as C code and assembly (\hyperref[subsubsec:static_linking]{E5} and \hyperref[subsubsec:assembly]{E6}) (also reported in security-critical projects like Kubernetes and Go-Ethereum) suggests that developers may trade the built-in security mechanisms in Go for less secure code, growing the attack surface. Although less common, the use of plugins (\hyperref[subsubsec:dynamic_linking]{E7}) is also present. Here, the loading and execution of externally pre-built code requires thorough security analysis.

Now, let us consider Listing \ref{lst:attack_example}, demonstrating a real-world example of a dynamic external execution attack vector (E8) discovered by \toolname{} in a Kubernetes dependency \href{github.com/mistifyio/go-zfs/}{\texttt{mistifyio/go-zfs}}. This small library implements a wrapper for ZFS command line tools and is maintained by 23 common GitHub users with no affiliation to the Kubernetes project. 

It is important to clarify that the reported example is not a command injection vulnerability exploitable through a user-controlled input, but rather a potential attack vector, according to the definition in Section \ref{sec:attacks}. It would be a good place for attackers to hide code in order to perform a supply chain attack. 
For example, a malicious actor would compromise a repository (e.g., a pull request that introduces a hidden payload or alters code in a subtle way), altering the dependency control flow to taint the arguments passed to \texttt{Run} (line 1), enabling execution of custom and lower-level OS commands. Here, the use of \texttt{exec.Command} presents a risk of a supply chain compromise, potentially affecting all the dependents of the \texttt{go-zfs} library, including Kubernetes.

\begin{lstlisting}[language=go,style=go,caption={Real-world attack vector.},label={lst:attack_example}]
func (c *command) Run(arg ...string) ([][]string, error) {
	cmd := exec.Command(c.Command, arg...)
        ...
}
\end{lstlisting}

For attackers, it is tempting to target widely-used dependencies to maximize the impact on downstream applications. In addition, dependencies with a larger attack surface are more likely to be targeted because they provide more opportunities for hiding the execution of malicious code. Therefore, these dependencies require closer scrutiny.

To sum up, \toolname{}'s results validated the applicability of our taxonomy. In Sec. \ref{sec:attacks} we demonstrated the feasibility of each attack vector developing PoCs for them. In this section, we confirm the increased security risks associated with all 12 vectors by showing their prevalence in real-world Go projects. This effectively motivates the need for additional scrutiny and prioritization of these vectors during security audits.

\subsection{Attack Surface over Versions}
\label{subsubsec:diff_analysis}
The attack surface of a software package is dynamic, with new attack vectors potentially introduced with each new version release. \toolname{} can be leveraged to continuously monitor these changes, by analyzing and comparing multiple versions of a package. 
In this experiment, we aim to demonstrate the dynamic nature of the attack surface and highlight the importance of regularly auditing dependencies for emerging threats, reasoning the \toolname{}'s use case described in \ref{subsubsec:updates}.

\paragraph{Methodology} To examine how the attack surface evolves in an actively developed project and understand the impact of software updates on supply chain security, we used \toolname{} for differential analysis. We evaluated the last five major releases of the popular Kubernetes project (versions 1.26 through 1.30) to investigate the prevalence of different attack vectors as the codebase evolved. The selected releases cover a time window of around 16 months.

\paragraph{Results} The results obtained from running \toolname{} on five different versions of Kubernetes are shown in Table \ref{table:comparison}. The reported occurrences suggest that the attack surface is not static and can significantly vary as new features are added, and code is refactored over release cycles. 

For example, the introduction of \textit{structured authorization configuration} likely contributed to the increase in interfaces (\hyperref[subsubsec:interfaces]{E3}) from Kubernetes v1.28 to v1.29 \cite{k8s_changelog}, as this feature allows for an extensible and decoupled authorization mechanism. Similarly, the \textit{in-place update of pod resources}, introduced in the same upgrade, potentially contributed to the increased usage of reflection (\hyperref[subsubsec:reflection]{E2}), as it might require dynamically accessing pod structures.

On the other hand, it is reasonable to observe a decrease in the attack vectors between Kubernetes versions 1.29 and 1.30, as the codebase has reduced in size. This reduction can often be attributed to the stabilization of features and the removal of redundant code, suggesting that refactoring can reduce significantly the attack surface for Go packages.

\begin{table}[t!]
\caption{Vector usage in 5 different releases of Kubernetes}
\label{table:comparison}
\small
\begin{tabular}{lrrrrr}
\toprule
 & v1.26 & v1.27 & v1.28 & v1.29 & v1.30 \\
\midrule
\hyperref[subsubsec:go_generate]{P1} & 106 & 95 & 104 & 147 & 119 \\
\hyperref[subsubsec:testing]{P2} & 10,067 & 10,291 & 10,501 & 10,770 & 10,339 \\
\hyperref[subsubsec:global_variable]{I1} & 35 & 35 & 36 & 37 & 36 \\
\hyperref[subsubsec:init]{I2} & 8319 & 8,304 & 8,320 & 8,336 & 1,108 \\
\hyperref[subsubsec:constructor]{E1} & 43,701 & 43,102 & 43,822 & 44,389 & 38813 \\
\hyperref[subsubsec:reflection]{E2} & 1,543 & 1,526 & 1,540 & 1,568 & 1,528 \\
\hyperref[subsubsec:interfaces]{E3} & 93,076 & 87,700 & 89,849 & 90,901 & 85277 \\
\hyperref[subsubsec:unsafe]{E4} & 7,769 & 7,744 & 7,883 & 7,865 & 8,105 \\
\hyperref[subsubsec:static_linking]{E5} & 820 & 792 & 795 & 797 & 803 \\
\hyperref[subsubsec:assembly]{E6} & 1,495 & 1,494 & 1,494 & 1,495 & 1,495 \\
\hyperref[subsubsec:dynamic_linking]{E7} & 1 & 1 & 1 & 1 & 1 \\
\hyperref[subsubsec:dynamic_execution]{E8} & 242 & 236 & 236 & 244 & 230 \\
\midrule
LOC & 3,967,186 & 3,855,352 & 3,934,131 & 4,004,993 & 3,728,835 \\
\bottomrule
\end{tabular}
\end{table}

In conclusion, \toolname{} is a tool for auditors to continuously monitor the attack surface of dependencies. As software dependencies are updated over time, new attack vectors may be introduced, potentially increasing the risk for downstream consumers. Continuously monitoring these changes is crucial for maintaining a secure software supply chain.

\section{Related Work}
\label{sec:related}

\paragraph{Generic Supply Chain Attack Taxonomies} Taxonomies aim to categorize and classify threats, establishing comprehensive knowledge for software supply chains. 

Ohm et al. \cite{ohm2020backstabber} collect and analyze 174 malicious real-world packages exploited in supply chain attacks for JavaScript, Java, Python, PHP, and Ruby packages. The analysis grounds two taxonomies. The first taxonomy classifies 18 attack methods for injecting malicious code into a dependency tree. The second one classifies 8 vectors for executing the injected malicious code, belonging to one of two branches: a software lifecycle phase (install scripts, test case, and runtime), or a conditional trigger. The second taxonomy classifying execution vectors is analogous to ours. However, they do not consider any Go knowledge and only provide a general taxonomy, without concrete attacker methods. Our taxonomy is the first one specialized for Go, based on deep analysis of Go-specific triggers within the Go-specific package lifecycle. 

Ladisa et al. \cite{ladisa2023sok} extend the Ohm et al. taxonomy for malicious code \textit{injection} to 107 attack vectors. They survey grey and white literature, considering exploited and non-exploited attacks, but do not categorize execution strategies for malicious code after injection. Conversely, our taxonomy focuses on the \textit{execution} strategies of already injected malicious code, in the context of Go. 

In another work, Ladisa et al. \cite{ladisa2023hitchhiker}  consider language and package manager features that can be exploited for executing malicious code. They find 7 general execution techniques, divided into 3 install-time, and 4 runtime methods. They map each technique's applicability in 7 languages, including Go.
This taxonomy over the execution of malicious code is related to ours. However, they only identify three runtime vectors for Go, missing 9 possible attack vectors. Our work proposes a novel and useful taxonomy of 12 Go-specific attack vectors throughout 3 package lifecycle phases. Most importantly, we implement a prototype tool, \toolname{}, which is able to find all instances of attack vectors from our proposed taxonomy.

\vspace{-7pt}
\paragraph{Go Security Tools}
There are security tools for Go that focus on static analysis of dependencies.

\textit{Go vet} \cite{govet} reports code quality and correctness issues such as unused variables and potential race conditions. However, it only identifies unsafe pointer conversions (Sec. \ref{subsubsec:unsafe}) as an attack vector and does not detect any other vectors that \toolname{} does. \textit{Gosec} \cite{gosec} identifies violations of 36 security rules (e.g., URL injection in HTTP requests and path traversal vulnerabilities). However, it does not report on any of the 12 attack vectors defined in our taxonomy. 
\textit{Go AST} \cite{goast} provide a generic \textit{Rego} policy engine, that allows users to create customized analysis rules. This tool can be used to identify simple attack vectors but does not suffice for more complex analysis, like the ones for \textit{assembly} (\hyperref[subsubsec:assembly]{E6}) and interfaces (\hyperref[subsubsec:interfaces]{E3}) that \toolname{} supports. 

\textit{Govulncheck} \cite{govulncheck} and \textit{Nancy} \cite{nancy} are vulnerability scanning tools that report the occurrences of vulnerable dependencies within a project. While Govulncheck uses \href{https://pkg.go.dev/vuln/}{Go Vulnerability Databases} for vulnerability matching, Nancy is powered by \href{https://ossindex.sonatype.org/}{Sonatype OSS Index}. In contrast, \toolname{} does not report known vulnerabilities based on a database, but rather detects the attack surface of a Go module according to its attack vector taxonomy. 

\textit{Capslock} \cite{capslock} is a capability analysis tool reporting privileged operations in a given package and its dependencies. The tool combines analysis of the AST and SSA (static single assignment) representations to compute the call graph for the Go program. By analyzing the call graph, Capslock identifies capabilities such as network and file system access. In addition, Capslock identifies reflection, CGO, plugins, and external execution, that align with four of our attack vectors. \toolname{} complements Capslock by providing more information about the attack surface of a Go module according to the taxonomy. 

\section{Conclusion}
\label{sec:conclusion}
Software supply chain attacks targeting open-source dependencies pose a significant threat. We propose a novel taxonomy categorizing 12 distinct attack vectors for the Go ecosystem, spanning the pre-build, initialization, and execution phases of the Go package lifecycle. To aid in identifying these vectors, we developed \toolname{}, a static analysis tool that analyzes Go modules and reports occurrences of the defined vectors. Our evaluation demonstrates the applicability of the taxonomy on real-world Go projects and the effectiveness of \toolname{} in highlighting potential supply chain risks. By raising awareness and providing tools like \toolname{}, we aim to empower the community to proactively address these risks and foster a more secure software supply chain in the Go ecosystem.

\begin{acks}
This work was partially supported by the Wallenberg Artificial Intelligence, Autonomous Systems and Software Program (WASP) funded by Knut and Alice Wallenberg Foundation, by the Swedish Foundation for Strategic Research (SSF), by the GENIO project (CUP B69J23005770005) funded by MIMIT, "Accordi per l'Innovazione" program and by MUR PRIN 2022, project FLEGREA, CUP E53D23007950001 (\url{https://flegrea.github.io}). Some computation was enabled by resources provided by the National Academic Infrastructure for Supercomputing in Sweden (NAISS).

\end{acks}


\balance
\bibliographystyle{ACM-Reference-Format}
\bibliography{bibfile}

\end{document}